\documentclass[aps,prl,twocolumn,floatfix]{revtex4}
\usepackage{amsmath}
\usepackage{graphicx}

\begin{document}

\title{Making Sense of the Legendre Transform}
\author{R. K. P. Zia$^1$, Edward F. Redish$^2$, and Susan R. McKay$^3$} 

\date{January 30, 2009}

\begin{abstract}
The Legendre transform is an important tool in theoretical physics, playing
a critical role in classical mechanics, statistical mechanics, and
thermodynamics. Yet, in typical undergraduate or graduate courses, the power of
motivation and elegance of the method are often missing, unlike the treatments
frequently enjoyed by Fourier transforms. We review and modify the
presentation of Legendre transforms in a way that explicates the formal
mathematics, resulting in manifestly symmetric equations, thereby clarifying
the structure of the transform algebraically and geometrically. Then we
bring in the physics to motivate the transform as a way of choosing
independent variables that are more easily controlled. We demonstrate how the
Legendre transform arises naturally from statistical mechanics and show how
the use of dimensionless thermodynamic potentials leads to more natural and
symmetric relations.
\end{abstract}

\affiliation{$^1$Department of Physics, Virginia Polytechnic Institute and State University, Blacksburg, VA 24061 USA}
\affiliation{$^2$Department of Physics, University of Maryland, College Park, MD 20742 USA}
\affiliation{$^3$Department of Physics and Astronomy, University of Maine, Orono, ME, 04469 USA}

\maketitle

\section{Introduction}

The Legendre transform is commonly used in upper division and graduate
physics courses, especially in classical mechanics,\cite{CCC} statistical
mechanics, and thermodynamics.\cite{SM,JWC} Most physics majors are first
exposed to the Legendre transform in classical mechanics, where it provides
the connection between the Lagrangian $\mathcal{L}(\dot{q})$ and the
Hamiltonian $\mathcal{H}(p)$, and then in statistical mechanics where it
yields relations between the internal energy $E$ and the various
thermodynamic potentials. Despite its common use, the Legendre transform
often appears in an ad hoc fashion, without being presented as a general and
powerful mathematical tool in the way the Fourier transform is.

In this paper we present a pedagogical introduction to the Legendre
transform, discuss it as a mathematical process, and display some of its
general properties. Since some students prefer algebraic approaches and some
prefer geometric ones, we discuss the transform from both points of view and
relate them. We then motivate the transform in terms related to physical
conditions and constraints. We emphasize some of the symmetries and
structures of the transform and present a series of increasingly complex
examples beginning with classical mechanics and going through examples in
statistical mechanics. We end with some remarks on more general versions of the Legendre transform, as well as other areas in which it is widely used. 

\section{The Legendre transform as an alternative way to display information}

In our experience, many students can manage the rules for generating a
Hamiltonian from a Lagrangian or switching between thermodynamic potentials
quite well, but express discomfort when asked about the Legendre transform
as a general mathematical tool. One possible reason is that in introductory
physics we often treat a function as a relation between physical rather than
mathematical quantities. Thus, when we are thinking about physical functions
we tend not to pay attention to the particular functional form the
mathematical function uses to encode physical information.\cite{EFR} For
example, if we are describing a position as a function of time, we might
write it as $x(t)$. We do not bother to change the symbol $x$ if we decide
to give $t$ in milliseconds instead of in seconds. If we write the
temperature as a function of position as $T(\vec{r})$, we do not change the
symbol if we switch to a different coordinate system or measuring scale. In
contrast, the Legendre transform is explicitly about how information is
coded in the functional form.

In addition, students are usually first introduced to the Legendre transform
as the transformation in classical mechanics from the Lagrangian to the
Hamiltonian. This transformation involves the switch from the velocity to
the momentum variable in the non-relativistic kinetic energy. In the context
of non-relativistic particle motion with velocity independent potentials,
the transform involves the kinetic energy, the most trivial function to
which the Legendre transform can be applied. The result looks like a shift
in units (from $v$ to $mv$ as an independent variable) so that it seems
pointless. Because the position variable $q$ plays no role in the transform
and typically appears only in $V$, the result is often regarded as a
mysterious change of the sign of $V$: 
$\mathcal{L}=T-V$ vs. $\mathcal{H}=T+V$.

In the rest of this section we motivate the Legendre transform as a general
mathematical transformation and describe a method that displays its general
properties and symmetries.

For clarity, we begin with a single variable $x$ and consider multivariate
functions later. Generally, a function expresses a relation between two
parameters: an independent variable or control parameter $x$ and a dependent
value $F$. This information is encoded in the functional form of $F(x)$. For
later convenience, we will also denote such a relationship or ``encoding''
as $\left\{ F,x\right\} $.

In some circumstances it is useful to encode the information contained in a
function $F(x)$ in a different way. Two common examples are the Fourier
transform and the Laplace transform. These transforms express the function $%
F $ as sums of (complex or real) exponentials, and display the information
in $F$ in terms of the amount of each component contained in the function
rather than in terms of the value of the function. In the notation
introduced above, $\left\{ \tilde{F},k\right\} $ encodes the same
information as $\left\{ F,x\right\} $. For the Fourier transform, 
$\tilde{F} \left( k\right) \equiv \int e^{ikx}F\left( x\right) dx$ 
is an explicit ``transformation'' between the two encodings.

Given an $F(x)$, the Legendre transform provides a more convenient way of
encoding the information in the function when two conditions are met: (1)
The function (or its negative) is strictly convex (second derivative always positive) and smooth (existence of ``enough'' continuous derivatives). (2)
It is easier to measure, control, or think about the derivative of $F$ with
respect to $x$ than it is to measure or think about $x$ itself.

Because of condition (1), the derivative of $F(x)$ with respect to $x$ can
serve as a stand in for $x$; that is, there is a one-to-one mapping between 
$x$ and $dF/dx$. (We remark on relaxing this condition in the last section.) 
The Legendre transform shows how to create a function that contains the same information as $F(x)$ but as a function of $dF/dx$.

\section{The mathematics of the Legendre transform}

We first consider a single, smooth convex function of a single variable.
There are many equivalent ways to characterize convex functions. The most
convenient one is that the second derivative $d^2F(x)/dx^2$ is always
positive. A second characterization of our condition is that the slope
function 
\begin{equation}
s(x) \equiv \frac{dF(x)}{dx}  \label{s-def}
\end{equation}
is a strictly monotonic function of $x$ (since this also permits us to treat
functions whose negative is convex).

A graphical way to see how the value of $x$ and the slope of a convex
function can stand in for each other can be seen by considering the example
in Fig.~1, where the curve drawn to represent $F$ is convex. As we move
along the curve to the right (as $x$ increases), the slope of the tangent to
the curve continually increases. In other words, if we were to graph the
slope as a function of $x$, it would be a smoothly increasing curve, such as the example in Fig.~2. If the second derivative $\frac{d^2F(x)}{dx^2} $ exists
(everywhere within the range of $x$ in which $F$ is defined; part of the
condition for a smooth $F$), there is a unique value of the slope for each
value of $x$, and vice versa. The corresponding mathematical language is
that there is a 1 to 1 relation between $s$ and $x$; that is, the function $%
s(x)$ is single-valued and can be inverted to give a single-valued function $%
x(s)$.

\begin{figure}[t]
\begin{center}
\resizebox{85mm}{!}{\includegraphics{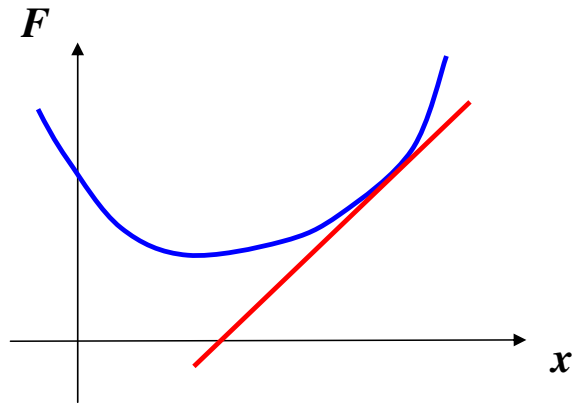}}
\end{center}
\par
\vspace{-0.7cm} 
\caption{The graph (blue online) of a convex function$F\left( x\right) $.
The tangent line at one point is illustrated (red online).}
\label{fig:Fig1}
\end{figure}

\begin{figure}[t]
\begin{center}
\resizebox{85mm}{!}{\includegraphics{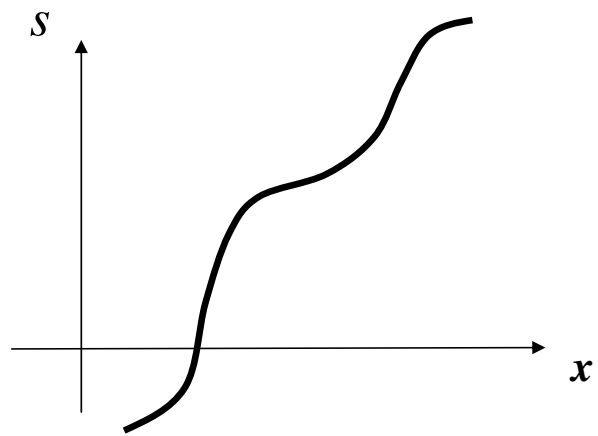}}
\end{center}
\par
\vspace{-0.7cm} 
\caption{The graph of $s\left( x\right)$, the slope of a convex function.}
\label{fig:Fig2}
\end{figure}

In this way, we could then start with $s$ as the independent variable, use
the inverse function to get an unique value of $x$, and then insert that
into $F(x)$ to access $F$ as a function of $s$. The standard notation for
such a function is $F(x(s))$. If we insist on a new encoding of the
information in $F$ (in terms of $s$ instead of $x$), this straightforward
``function of a function approach'' would appear to be the most natural way.

Instead, the Legendre transform of $F(x)$ is defined quite differently, and
seemingly quite unnatural: 
\begin{equation}
G(s)=sx(s)-F(x(s))\,\,.  \label{G-def}
\end{equation}
Typically, this formula is presented with little motivation or explanation,
and leaves the students to ponder: Why? Why the extra $sx$? Why the minus
sign? Frequently, the instructor or the author (of a textbook) invokes
another magical relation to answer such queries. Only with this peculiar
definition can we have the property that ``the slope of $G(s)$ is just $x$%
'': 
\begin{equation}
x(s)=\frac{dG}{ds}\,\,.  \label{G-dot}
\end{equation}
This result also requires a careful calculation.

\subsection{A graphic-geometric approach}

Before providing ways to appreciate this definition of the Legendre
transform, as well as how never to forget ``which sign goes where,'' we
present a graphical route to the transform. Consider the plot of $F$ versus $%
x$ in Fig.~3. Choose a value of $x$, which is represented by the length of
the horizontal line labeled by $x$. Go up to the value on the function
curve, $F(x)$. This value corresponds to the length of the vertical line
labeled by $F$. Next, draw the tangent to the curve at that point. The slope
here is labeled $s$, as emphasized by the call out bubble. Extend this
tangent until it hits the ordinate (the ``$F$ axis''). In this example, the
intercept is negative and is labeled as $-G$, with a positive $G$. This
value corresponds to the length of the thick vertical line labeled by $G$.
This length is reproduced (thin line) just below the line labeled $F$.
Because the slope of the tangent is $s$, the length of the dotted vertical line
is $sx$. From this picture, it is quite clear that $sx=F+G$. In this light, 
the peculiar definition of the Legendre transform in Eq.~(\ref{G-def}) 
appears natural. The minus sign in the definition is seen as a way of
retaining the symmetry and simplicity of the geometrical statement: ``In the
triangle, the slope (tangent) times the adjacent side equals the opposite
side, which is the sum of $F$ and $G$.''

\begin{figure}[t]
\begin{center}
\resizebox{85mm}{!}{\includegraphics{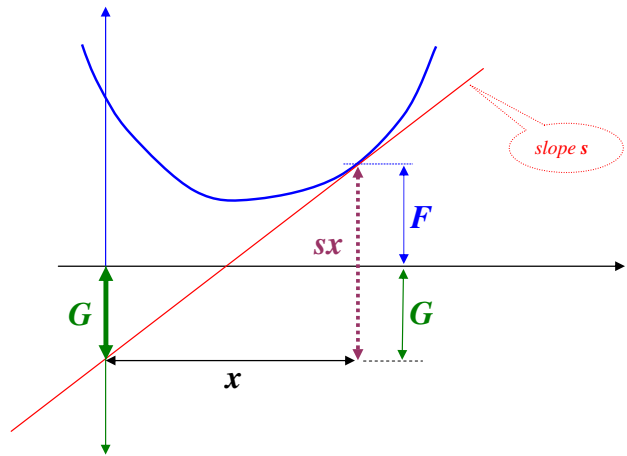}}
\end{center}
\par
\vspace{-0.7cm} 
\caption{Graphic representation of the Legendre transform, 
$G\left( s\right) $, of $F\left( x\right) $. See text for an 
explanation of various quantities (color online).}
\label{fig:Fig3}
\end{figure}

\subsection{Symmetric representation of the Legendre transform}

This symmetric, geometrical construction allow us to display a number of
useful and elegant relations that shed light on the workings of the Legendre
transform. In particular, we consider the symmetries associated with the
inverse Legendre transform, extreme values, and derivative relations.

Ordinarily, the inverse of a transformation is distinct from the transform
itself. For example, an inverse Laplace transform is not given by the same
formula. The Legendre transform distinguishes itself in that it is its own
inverse. In this sense, it resembles (geometric) duality transformations.
Symbolically, we may denote this relationship as: 
\begin{equation}
\left\{ F,x\right\} \Leftrightarrow \left\{ G,s\right\} \,\,.  \label{FxGs}
\end{equation}
Specifically, if we perform the Legendre transform a second time, we recover
the original function. (If the restriction of convexity is relaxed, this 
statement must be revised, as remarked in the final section.) In other words, suppose we start with the 
function $G(s)$ and calculate its Legendre transform. Of course, as we will
see, $G(s)$ satisfies our conditions: convex and smooth. So, we start with 
\begin{equation}
y(s)=\frac{dG}{ds}  \label{y=G-dot}
\end{equation}
and invert the monotonic function $y(s)$ to $s(y)$. Next, we construct 
\begin{equation}
H(y)=ys(y)-G(s(y))\,\,,
\end{equation}
which can be rewritten as 
\begin{equation}
G=sy-H\,\,.  \label{eq:this1}
\end{equation}
If we compare this equation and Eqs.~(\ref{G-def}), we see that we can 
identify $\{H,y\}$ with $\{F,x\}$. Thus, the
Legendre transform of $G$ is the original function $F$, leading to the
statement: the Legendre transform is its own inverse. This ``duality'' of
the Legendre transform, shown symbolically in Eq.~(\ref{FxGs}), is best
displayed by the symmetric form 
\begin{equation}
G(s)+F(x)=sx\,\,.  \label{F+G}
\end{equation}
This equation~should be read carefully. Despite its appearance, there is
only one independent variable: either $s$ or $x$. Referred to as a conjugate
pair, these two variables are related to each other, through either 
$x(s)= dG(s)/ds$ or $s(x)=dF(x)/dx$. A careful
writing of Eq.~(\ref{F+G}) would read either $G(s)+F(x(s))=sx(s)$ or 
$G(s(x))+F(x)=s(x)x$. To double check the consistency with 
Eqs.~(\ref{s-def}) and (\ref{G-dot}), we can start with, say, the first 
of these equations and differentiate with respect to $s$. Applying the
chain rule for $dF/ds=(dF/dx)(dx/ds)$, we recover $dG/ds=x$.

\subsection{Properties of the extrema}

The example in Fig.~3 shows a convex function $F(x)$ with a unique minimum.
Let us denote this point by $F_{\min }=F(x_{\min })$. The slope of the
tangent vanishes here, that is, $s(x_{\min })=0$. If we substitute this
point into Eq.~(\ref{G-def}), we find that the minimum value of $F$ is 
\begin{equation}
F_{\min }=-G(0)\,\,.  \label{Fmin}
\end{equation}
It is straightforward to show that a ``dual'' relation exists, namely, the
minimum value of $G$ is $G_{\min }=-F(0).$ To appreciate the geometric
meaning of this equation, we need only to inspect Fig.~3 and see that $-G$,
the $y$-intercept of the tangent to the curve $F(x)$, never reaches beyond 
$F(0)$.

Exploiting Eq. (\ref{F+G}), both this special example and the case of
general extrema can be cast in an ``easy-to-remember'' symmetric form.
Suppose $F$ takes on its extremal value at $x_{\mathrm{ext}}$, then we have
a horizontal tangent line and by definition, $s(x_{\mathrm{ext}})=0$.
Similarly, if $G$ is at its extremum at $s_{\mathrm{ext}}$, we have 
$x(s_{\mathrm{ext}})=0$ due to Eq.~(\ref{G-dot}). In either case, 
the right hand side of Eq. (\ref{F+G}) vanishes and we have 
\begin{equation}
G(0)+F(x_{\mathrm{ext}})=0\quad \ \text{and}\ \quad G(s_{\mathrm{ext}%
})+F(0)=0\,\,.  \label{eq:this2}
\end{equation}

\subsection{Symmetric representation of the higher derivatives}

Since the Legendre transform is a ``dual'' relationship, we can expect 
manifestly symmetric relations beyond the ones we have seen so far:
\begin{equation}
G(s)+F(x)=sx  \label{F+G}
\end{equation}
and 
\begin{equation}
\frac{dG}{ds}=x\quad \ \text{and}\ \quad \frac{dF}{dx}=s\,\,.  \label{dG+dF}
\end{equation}
From these, we can obtain an infinite set of relations linking $G$ and $F$ by
taking derivatives of $G+F=sx$ with respect to $s$ or $x$. 
Because each function depends on only one
variable, the differentials can be easily identified. Thus, differentiating
the equations in (\ref{dG+dF}) with respect to $s$ or $x$ as appropriate, we
find 
\begin{equation}
\frac{d^2G}{ds^2}=\frac{dx}{ds}\quad \ \text{and}\ \quad \frac{d^2F}{dx^2}=%
\frac{ds}{dx}\,\,.
\end{equation}
But $\left( dx/ds\right) (ds/dx)=1$, so we have 
\begin{equation}
\left( \frac{d^2G}{ds^2}\right) \left( \frac{d^2F}{dx^2}\right) =1\,\,.
\label{d2}
\end{equation}
Let us emphasize once more that the variable $s$ in the first factor and the 
$x$ in the second are not independent, but linked through Eqs. (\ref{dG+dF})!

Equation~(\ref{d2}) illustrates the importance of (strict) convexity so that
neither derivative ever vanishes. An interesting result is that the local
curvatures of the Legendre transforms are inverses of each other in a manner
reminiscent of the uncertainty relation $\Delta x\Delta k\approx 1$. For
simplicity, suppose $F$ is dimensionless but $x$ is not,\cite{note4} so that 
$s$ has the dimension of $1/x$. With this convention it is easy to check the
units of Eqs.~(\ref{F+G}, \ref{dG+dF}, and \ref{d2}).

If we differentiate Eq.~(\ref{d2}) again, we can write a symmetric relation
for the third derivative: 
\begin{equation}
\frac{d^3G}{ds^3}\left[ \frac{d^2G}{ds^2}\right] ^{-3/2}+\frac{d^3F}{dx^3}%
\left[ \frac{d^2F}{dx^2}\right] ^{-3/2}=0\,\,.  \label{d3}
\end{equation}
Notice that each term is again dimensionless, since the units of the various
derivatives precisely cancel.

It is possible to derive an infinite set of such relations for higher
derivatives by differentiating further. Such an exercise also shows that if
$F$ is smooth (with a well defined $n^{th}$ derivative), then so is $G$. The
relations for higher derivatives do not have forms as simple as Eqs. (\ref
{F+G}), (\ref{dG+dF}), (\ref{d2}) and (\ref{d3}), but become more and more
 complex.

\section{Examples of the Legendre transform in single-particle mechanics}

It is useful to provide some physical examples to illustrate these
relations. The simplest is a quadratic function $F(x)=\alpha x^2/2$.
For this function we easily find that $s=\alpha x$ and $x=s/\alpha $,
leading to $G(s)=s^2/2\alpha $. The curvatures in $F$ and $G$ ($\alpha $ and 
$1/\alpha $, respectively) are inverses of each other as required by Eq.~(%
\ref{d2}). All derivative relations beyond this level are trivial, i.e., $%
0=0 $.

This example corresponds to a single non-relativistic particle with mass $m$
moving in an external potential $V(q)$. The Legendre transform connects the
Lagrangian $\mathcal{L}(\dot{q})$ to the Hamiltonian $\mathcal{H}(p)$. Only
the kinetic term, which depends on $\dot{q}$ or $p$ is affected by the
transform, as the potential depends on an entirely different variable: $q$.
There, $x\to \dot{q}$, $F\to \mathcal{L}$, $\alpha \to m$, $s\to p$, and $%
G\to \mathcal{H}$, so that $\mathcal{L}=m\dot{q}^2/2\Leftrightarrow \mathcal{%
H}=p^2/2m$. However, since $V(q)$ is just a ``spectator'' in the Legendre
transform, it must appear with opposite signs in $F$ and $G$ (i.e., $%
\mathcal{L}$ and $\mathcal{H}$), in order to satisfy $F+G=sx$ (i.e., $%
\mathcal{L}+\mathcal{H}=p\dot{q}$, with no $q$ anywhere). Thus, we see the 
origin of the mysterious sign change in $V$ when we go from the Hamiltonian
to the Lagrangian.

Relativistic kinetic energy is a more interesting case. Here, we go the
other way and start with momentum and generate a velocity as the slope of
the function. The relativistic kinetic energy as a function of momentum is $%
\mathcal{H}(p)=\sqrt{p^2+m^2}$ (with $c\equiv 1$). $\mathcal{H}(p)$ is
convex and its slope at a point $p$ is 
\begin{equation}
v \equiv \frac{d\mathcal{H}}{dp}=\frac p{\sqrt{p^2+m^2}}\,\,,
\end{equation}
giving the familiar result 
\begin{equation}
p=mv /\sqrt{1-v^2}\,\,.
\end{equation}
Creating the Legendre transform using this pair of variables leads to the
Lagrangian\cite{JDJ} 
\begin{equation}
\mathcal{L}(v)=pv -\mathcal{H}(p(v))=-m\sqrt{1-v^2}\,\,.
\end{equation}
This example can also be written in terms of the function $F(x)=\cosh
\lambda x$. The demonstration is left to the reader. (Hint: see Ref.~%
\onlinecite{TW}.)

Let us turn to a less familiar example, one that is so trivial that it does
not appear in typical textbooks. Yet it sets the stage for examining the
role of the Legendre transform in equilibrium statistical mechanics.

Consider a particle in a one-dimensional convex potential well, $U(x)$,
which has a unique minimum at $x_{\min }$. An example would be a particle
attached to a wall by a non-ideal spring, with $x$ being the distance from
the point where the coils of the spring are fully compressed. The potential
is effectively infinite at $x=0$, decreases to a minimum at its natural
extension, and then increases for larger $x$. (We restrict our attention to
positive values of $x$, but less than the breaking point of the spring.) 
Another example of $U$ is the potential that binds two atoms into a 
molecule (though such $U$'s are rarely convex for all separations).

The particle is stationary only if it is at $x_{\min }$ for all time. If it
is subjected to an additional external applied force $f$, then it will reach
a new stationary point $x_0$, which is the solution to the equation 
\begin{equation}
\left. \frac{dU}{dx}\right| _{x_0}=f\,\,.  \label{x0}
\end{equation}
To emphasize the dependence of this point on $f$, we write $x_0(f)$. We can
ask the inverse question: If we want the particle to settle at $x_1\neq
x_{\min }$, what force do we need to apply? The answer is $f(x_1)$, a force
that depends on which $x$ we choose. A little thought leads us to the
explicit functional form: $f(x_1)=dU/dx|_{x_1}$. There is nothing special
about the subscripts here and we may just as well write 
\begin{equation}
f(x)=\frac{dU}{dx}\,\,,  \label{f=dU/dx}
\end{equation}
and $x(f)$ instead of $x_0(f)$.

Although Eq.~(\ref{f=dU/dx}) gives $f(x)$ explicitly, we may ask if there is
a counterpart to $U$ which provides the inverse, $x(f)$, explicitly. If so,
we can simply plug $f$ into the expression and arrive at the new equilibrium
position. The answer is the Legendre transform of $U$, namely, 
$V(f)=fx-U(x(f))$. We leave it to the reader to show that 
\begin{equation}
x(f)=\frac{dV}{df}  \label{x=dV/df}
\end{equation}
is the companion to Eq.~(\ref{f=dU/dx}).

All the details can be worked out for the simple example of the mass on a
spring, $U(x) = kx^2/2$. This example is the analog of the non-relativistic
kinetic energy Legendre transform. The reader can easily demonstrate that
the Legendre transform equation $U + V = fx$ becomes $(f-kx)^2 = 0$,
yielding the relation between $f$ and the new equilibrium point $x$.

Note that the information about the system (for example,
wall-spring-particle complex) is fully contained in either $U$ or $V$. The
only difference is in the coding: $\left\{ U,x\right\} $ vs. $\left\{
V,f\right\} $. Although $U$ is the usual potential energy associated with
putting the particle at $x$, $V$ is a kind of potential associated with the
control $f$. In ordinary classical mechanics, such an approach seems
unnecessarily cumbersome for describing the simple problems we posed. Thus,
it is rightfully ignored in a course on classical mechanics. We include the
example here only as a stepping stone to the Legendre transform in
statistical mechanics and thermodynamics. There, multiple potentials are
essential.

\section{The Legendre transform in statistical thermodynamics}

The Legendre transform appears frequently in statistical thermodynamics when
different variables are ``traded'' for their conjugates. 
\cite{SM} Often, one of the variables is easy to think about while the other
is easy to control in physical situations.

The difficulty with making sense of the Legendre transform in thermodynamics
arises from two causes: (1) For historical reasons, Legendre transform
variables are not always chosen as conjugate pairs. (2) Many variables in
thermodynamics are not independent and are constrained by equations of
state, for example, $PV=Nk_BT$.

As an example of the first point, the conjugate to the total energy $E$ of a
system is the inverse temperature ($\beta =1/k_B T $). Yet, our daily
experience with the temperature $T$ is so pervasive that $T$ is used in most
of the relations. Thus, the familiar equation 
\begin{equation}
F=E-TS\,\,,
\end{equation}
which relates the Helmholtz free energy $F$ to the entropy $S$, obscures the
symmetry between $\beta $ and $E$, as well as the dimensionless nature of
the Legendre transform. In contrast, if we define the dimensionless
quantities 
\begin{equation}
\mathcal{S}\equiv S/k_B\ \text{and}\ \mathcal{F}\equiv \beta F\,\,,
\end{equation}
the ``duality'' between them can be beautifully expressed as 
\begin{equation}
\mathcal{F}(\beta )+\mathcal{S}(E)=\beta E\,\,.  \label{F+S}
\end{equation}

To elaborate the second point, we typically encounter a bewildering array of
thermodynamic functions (for example, entropy, Gibbs and Helmholtz free
energies, and enthalpy), a slew of variables (energy, temperature, volume,
and pressure), as well as a jumble of thermodynamic relations (with multiple
partial derivatives). In general, because of the multiple constrained
variables, none of these examples is as simple as those we have
considered, compounding the difficulty of both teaching and learning this
material.

Before discussing the generation of the standard thermodynamic potentials, we briefly summarize the basics of statistical mechanics. We will show how the
Legendre transform enters thermodynamics through the Laplace transform of 
partition functions in statistical mechanics. 

Equilibrium statistical mechanics is based on the hypothesis\cite{SM} that
for an isolated system, every allowed microstate is equally probable. The
high probability of finding a particular equilibrium macrostate is due to a
predominance of the number of microstates corresponding to that macrostate.
The classic example is a gas of $N$ identical, free, non-relativistic
structureless particles, confined in a $D$-dimensional box of volume $L^D$.
For this system a microstate is specified by the $2DN$ variables
corresponding to the positions and momenta of each particle, 
$\{\vec{r}_i,\vec{p}_i\}$, with $i=1,\ldots ,N$. Because the total energy 
$E$ is a constant for an isolated system, the fundamental hypothesis can be
represented as 
\begin{equation}
P\left( \{\vec{r}_i,\vec{p}_i\}\right) \propto \delta \left( E-\mathcal{H}(\{%
\vec{r}_i,\vec{p}_i\})\right) \,\,,
\end{equation}
where $P\left( \{\vec{r}_i,\vec{p}_i\}\right) $ is the probability of
finding the configuration of positions and momenta $\{\vec{r}_i,\vec{p}_i\}$
and $\mathcal{H}$ is the Hamiltonian. In this case $\mathcal{H}$ is
explicitly given by 
\begin{equation}
\mathcal{H}=\sum_ih(\vec{r}_i,\vec{p}_i)=\sum_i\left[ \frac{\vec{p}_i^2}{2m}%
+U(\vec{r}_i)\right] \,\,,
\end{equation}
where $m$ is the mass of each particle and $U$ is the confining potential,
which is zero for each component of $\vec{r} \in [0,L]$ and infinite otherwise.

The normalization factor for $P$ is 
\begin{equation}
\Omega (E)=\!\int_{r,p}\!\delta \left( E-\mathcal{H}(\{\vec{r}_i,\vec{p}%
_i\})\right) \,\,,  \label{omega}
\end{equation}
where the integral is over all $\{\vec{r}_i,\vec{p}_i\}$ from $-\infty $ to $%
\infty $. (The infinite values of $U$ restrict the actual position
integrations to the volume of the box.) We have also suppressed the other
variables that $\Omega $ depends on for now: $L$ and $m$. Note that $\Omega $ 
is just the volume of phase space available for our system and is also known
as the microcanonical partition function.

The standard approach evaluates the integral in Eq. (\ref{omega}) as
follows. The position integrals can be done explicitly because the only
dependence of the Hamiltonian on position is the confinement of the position
integrals to the allowed volume. These integrals yield a factor of $L^{ND}$.
The momentum integrals are done by computing the surface area of a sphere in 
$DN$ dimensions.

The entropy is introduced by the definition $S\equiv k_B\ln \Omega $. We
exploit the ``dimensionless entropy'' $\mathcal{S}$ and write 
\begin{equation}
\mathcal{S}(E)\equiv \ln \Omega (E)\,\,.  \label{SOmega}
\end{equation}
To proceed, we have two choices: the route that emphasizes the mathematics
or the physics.

\subsection{The route of mathematics}

Our task is straightforward: evaluate integrals with a constraint such as
Eq.~(\ref{omega}). Often, such integrals are not easy to perform. However,
exploiting the Laplace transform typically renders the integrand factorizable.
For example, the $DN$ integrations in Eq.~(\ref{omega}) become products of a
single integral. Specifically, we consider the Laplace transform of $\Omega
(E)$, 
\begin{equation}
Z(\beta )\equiv \!\int \!\Omega (E)e^{-\beta E}dE\,\,.  \label{Z}
\end{equation}
If we substitute Eq.~(\ref{omega}) for $\Omega (E)$, the delta function
permits us to do the $E$ integral giving 
\begin{equation}
Z(\beta )=\!\int_{r,p}\!e^{-\beta \mathcal{H}}\,\,.  \label{ZZ}
\end{equation}
 
Because $\mathcal{H}$ is a sum over the individual components, the integrand
factorizes and we have the result: 
\begin{equation}
\int_{r,p}e^{-\beta \mathcal{H}}=\!\int_{r,p}\prod_ie^{-\beta h(\vec{r}_i,%
\vec{p}_i)}=\left[ \int d\vec{r}d\vec{p}e^{-\beta h(\vec{r},\vec{p}%
)}\right] ^N.  \label{z^N}
\end{equation}
Being an integral in just $2D$ dimensions, the expression in $\left[
...\right] $ is much easier to handle. For the classic example above, the
integral is simply $L^D\left( 2\pi m/\beta \right) ^{D/2}$.
The attentive student will have noticed, from Eq. (\ref{ZZ}) that $Z$ is the canonical partition function and can appreciate the statement: The two partition functions are related to each other through a Laplace transform.

To return to our goal, $\Omega (E)$, we need to perform an inverse Laplace
transform, that is, 
\begin{equation}
\Omega (E)=\!\int_{\mathcal{C}}\!Z(\beta )e^{\beta E}d\beta \,\,,
\end{equation}
where $\mathcal{C}$ is a contour in the complex $\beta $ plane (running
parallel to and to the right of the imaginary axis). We define 
\begin{equation}
\mathcal{F}(\beta )\equiv -\ln Z(\beta )\,\,,
\end{equation}
and write the integral as 
\begin{equation}
e^{\mathcal{S}(E)}=\!\int_{\mathcal{C}}e^{-\mathcal{F}(\beta )+\beta
E}d\beta \,\,.
\end{equation}
To continue it is necessary to inject some physics. In this case, we expect
to be considering many particles, that is, large $N$. From Eq.~(\ref{z^N}),
we have $\mathcal{F}\propto N$, leading us to expect that the range of $E$
of interest is also $O(N)$. The standard tool to evaluate integrals with
large exponentials as integrands is the saddle point (or steepest decent)
method. Thus, we seek the saddle point in $\beta $, defined by setting the
first derivative of $\beta E-\mathcal{F}(\beta )$ to zero: 
\begin{equation}
\left. \frac{d[\beta E-\mathcal{F}]}{d\beta }\right| _{\beta _0}=0\,\,.
\end{equation}
In other words, we have 
\begin{equation}
\left. \frac{d\mathcal{F}}{d\beta }\right| _{\beta _0}=E\,\,.  \label{E0}
\end{equation}
We emphasize that $\beta _0$ should be regarded as a function of $E$ here.

In this approach, the integral in Eq.~(\ref{Z}) is well approximated by
evaluating the integrand at the saddle point, so that 
\begin{equation}
\Omega (E)\cong \exp [\beta _0E-\mathcal{F}(\beta _0)]\,\,,
\end{equation}
or using Eq.~(\ref{SOmega}) 
\begin{equation}
\mathcal{S}(E)+\mathcal{F}(\beta _0)=\beta _0E\,\,,  \label{LT-SF}
\end{equation}
with the understanding that $\beta _0$ and $E$ are related through Eq.~(\ref
{E0}). There is nothing significant about the subscript on $\beta $ and Eq.~(%
\ref{LT-SF}) is identical to Eq.~(\ref{F+S}). In other words, $\mathcal{S}$
and $\mathcal{F}$ are Legendre transforms of each other. Thus, we see that
(for situations involving a large parameter, $N$ in this case) the Laplace
and Legendre transforms, Eqs.~(\ref{Z}) and (\ref{LT-SF}) respectively, are
intimately related to each other as a result of the thermodynamic limit.

\subsection{The route of physics: interpretation of the equilibrium condition
}

Under what conditions does the internal energy move from one object to
another and under what conditions can it be changed to work? Part of the
answer lies in understanding which way the energy will move if we bring two
different systems into thermal contact. Why does it not go always from the
system with more energy to the one with less? Considering this question
leads us back to the Legendre transform.

When two systems (not necessarily of the same size or energy) are brought 
together and the combined system isolated, $E_{\mathrm{tot}}\equiv E_1+E_2$
will remain a constant and can be regarded as the control parameter.
The individual $E_j$'s are not fixed, and we ask the question: 
Starting at some initial values, how do they wind up at the final 
equilibrium partition $\{E_1^{*},E_2^{*}\}$? The answer lies with 
$\mathcal{S}_{\mathrm{tot}}(E_{\mathrm{tot}}|E_1,E_2)$, the entropy of the
combined system subjected to the specific partition of $E_{\mathrm{tot}}$ 
into $\{E_1,E_2\}$. The idea is $e^{\mathcal{S}_{\mathrm{tot}}}$ counts 
the number of allowed microstates associated with a particular partition
and so, carries the information of how probable that partition is. 
In general, calculating this quantity is not trivial. However, if we
focus on systems with extensive entropies, then we may write to a good
approximation: $\mathcal{S}_{\mathrm{tot}}=\mathcal{S}_1+\mathcal{S}_2$ with 
$\mathcal{S}_1=\mathcal{S}_1(E_1)$ and $\mathcal{S}_2=\mathcal{S}_2(E_2)$.
These statements are not trivial: We are injecting the physics that, under
the conditions specified, the entropies of each system do not depend on the
energy of the other.

Given these assumptions we can ask for what partition will $\mathcal{S}_{%
\mathrm{tot}}$ be maximum, or equivalently, which partition is the most
probable? If we write $E_2=E_{\mathrm{tot}}-E_1$ and recall that $E_{\mathrm{%
tot}}$ is fixed, this task is easy. The maximum occurs at $E_1^{*}$, defined
by 
\begin{equation}
\frac{d\mathcal{S}_{\mathrm{tot}}}{dE_1}|_{E_1^{*}}=0\,\,.
\end{equation}
But $dE_1=-dE_2$, so that we have 
\begin{equation}
\frac{d\mathcal{S}_1}{dE_1}|_{E_1^{*}}=\frac{d\mathcal{S}_2}{dE_2}%
|_{E_2^{*}}\,\,,  \label{eq}
\end{equation}
where $E_2^{*}=E_{\mathrm{tot}}-E_1^{*}$. This result is significant because
each side does not depend on the parameters of the other system. Thus, if we
associate a quantity with $d\mathcal{S}/dE$, which we define by 
\begin{equation}
\beta (E)\equiv \frac{d\mathcal{S}}{dE}\,\,,  \label{beta}
\end{equation}
then Eq.~(\ref{eq}) becomes 
\begin{equation}
\beta _1(E_1^{*})=\beta _2(E_2^{*})\,\,.
\end{equation}
In other words, the most probable partition occurs when the $\beta $ of one
system equals the $\beta $ of the other. Note that this condition does not
depend on the details of the two systems, such as composition, size, or
state (gas, liquid, solid, etc.). When the two systems are brought into
contact, energy will flow between them until they settle at values given by
this condition: the equality of a quantity, $\beta \equiv d\mathcal{S}/dE$,
associated with each of them separately. It is natural, therefore, to use
this quantity for describing our daily experience, namely, two systems, one
hot and one cold, will equilibrate at a common ``temperature'' ($T$) when
brought in contact with each other. Historically, many arbitrary scales were
used for $T$. Their relationships with the more natural quantity $\beta $
were not clarified later.

Besides providing a natural scale to describe ``hot'' and ``cold,'' can this
variable $\beta $ be exploited further? For a given system, we can write $%
\mathcal{S}(E(\beta ))$, but is that useful? The answer is connected to the
canonical ensemble, the (Helmholtz) free energy, and the Legendre transform
of $\mathcal{S}$. There is no need to reproduce here the standard derivation
of this ensemble and the Boltzmann factor $e^{-\beta \mathcal{H}}$. In the
previous subsection, we have already discussed the transformation between
the partition functions $Z(\beta )$ and $\Omega (E)$ and the relationship to
the Legendre transform between $\mathcal{S} (E)$ and $\mathcal{F}(\beta )$.

\subsection{How does the Legendre transform enter into thermodynamics?}

For convenience we summarize the key relations using dimensionless
potentials: 
\begin{align}
\Omega (E)=e^{\mathcal{S}(E)}\,\,,\,\,& 
Z(\beta )=e^{-\mathcal{F}(\beta)}\,\,, \\
\frac{d\mathcal{S}}{dE}=\beta \,\,,\,\,& 
\frac{d\mathcal{F}}{d\beta }=E\,\,,
\end{align}
where $Z\left( \beta \right) =\int dEe^{-\beta E}\Omega (E)$ and, in the
thermodynamic limit, $\mathcal{F}(\beta )+\mathcal{S}(E)=\beta E$. We can
now see where the Legendre transform enters and why it is useful. The
entropy $\mathcal{S}$ is a function of $E$, but the internal energy is
typically not easy to control. To put more energy into a system (or take
some out), we may give it some heat (or remove some). In other words, we
often manipulate $E$ by coupling the system to an appropriate thermal bath
and so, temperature (or $\beta $) becomes the ``control'' variable. In that
case, we can perform a Legendre transform of $\mathcal{S}\left( E\right) $
and work with $\mathcal{F}(\beta )$ instead. Since both $\left\{ \mathcal{S}%
,E\right\} $ and $\left\{ \mathcal{F},\beta \right\} $ contain the same
information about our system, it makes sense to deal with the more convenient
thermodynamic potential when we change the control on our system from
one variable to another.

Since the independent variable in a thermodynamic potential is to be
regarded as a control (or a constraint) parameter, the ``slope'' associated
with this function (e.g., $d\mathcal{S}/dE$, $d\mathcal{F}/d\beta $) carries
physically significant information, namely, the response of the system to
this control. The Legendre transform simply exchanges the role of the
variables associated with control and response. For the example discussed
above, it is physically easier to control $T$. It is also more familiar to
think of temperature (or $\beta $) as a control and the internal energy as
the response. Thus, the free energy $\mathcal{F}(\beta )$ is the more
appropriate potential, with $E=d\mathcal{F}/d\beta $ being the response. In
the transformed version, which is mathematically and conceptually easier to
grasp, $E$ is a constraint (conserved variable for an isolated system) and $%
\mathcal{S}\left( E\right) $ is the more appropriate potential. After we
understand the significance of its slope, $d\mathcal{S}/dE$, we can identify
the ``response'' $\beta $ with a measure for temperature. There are many
other examples of response/control pairs to which the same kind of
transformation may be applied, such as particle number and chemical
potential, polarizability and electric field, magnetization and magnetic
field, etc.

\section{Legendre transform with many variables}

The thermodynamic potentials depend on many variables other than just the
total energy $E$. Each variable that can be independently controlled elicits
a distinct response. As we construct Legendre transforms for each of these
control/response variable pairs, we generate a new potential. The result is
a plethora of thermodynamic functions. We again emphasize that all these
thermodynamic potentials carry the same information, but encoded in
different ways. We begin this section by discussing briefly the mathematical
structure of the multivariable Legendre transform and then apply it to
thermodynamics and statistical mechanics.

Consider the multivariate function $F(\vec{x})$, where $\vec{x}$ stands for $%
M$ independent variables: $x_1,\ldots ,x_M$. For convenience, suppose $F$ is
smooth and convex over all of this $M$-dimensional space. At every point $%
\vec{x}$, there will be $M$ slopes: 
\begin{equation}
s_m=\frac{\partial F}{\partial x_m}\equiv \partial _mF\,\,,
\end{equation}
and $M(M+1)/2$ second derivatives, $\partial _m\partial _\ell F$, which can
be regarded as a symmetric matrix. The convexity restriction requires that
all of the eigenvalues of this matrix are positive (or negative).\cite{CUR}
In the context of thermodynamics, convexity is the condition for stability
in equilibrium systems.\cite{note7} A standard corollary is that the
relation between $\{x_m\}$ and $\{s_m\}$ is 1 to 1, so that we can replace
any one of the $x_m$'s by the corresponding $s_m$ through a Legendre 
transform.

Because we can transform any number of the $x$'s, we may consider (up to) $%
2^M $ functions. For example, if we restrict ourselves to $(E,V)$ -- the
standard variables for the microcanonical ensemble of the ideal gas -- there
are four thermodynamic functions: entropy, enthalpy, Gibbs, and the
Helmholtz free energies. One way to picture the relation between so many
functions is to put them at the corners of an $M$-dimensional hypercube.
Each axis in this space is associated with a particular variable pair $%
(x_m,s_m)$. Going from one corner to an adjacent corner along a particular
edge corresponds to carrying out the Legendre transform for that pair. For
the $M=2$ example of $(x_1,x_2)=(E,V)$, the hypercube reduces to a square,
which is related, but not identical, to the square that appears in some
texts.\cite{SM,note8} Thanks to the commutativity of partial derivatives,
going from any corner to any other corner is a path independent process, so
that the function associated with each vertex is unique. For example, if we
exchange $(x_\ell ,x_m)$ for $(s_\ell ,s_m)$, the Legendre transform
relations would be the simple generalization of Eq.~(\ref{F+G}) 
\begin{widetext}
\begin{equation}
F(x_1,\ldots x_\ell ,\ldots x_m,\ldots x_M)+ 
G(x_1,\ldots s_\ell ,\ldots s_m,\ldots x_M) \\ 
= s_\ell x_\ell +s_mx_m\,\,,  \label{whatever}
\end{equation}
\end{widetext}
with\cite{note9} $\partial _\ell G=x_\ell $, $\partial _mG=x_m$, $\partial
_\ell F=s_\ell $, and $\partial _mF=s_m$. We should have given this $G$ some
special notation to denote that its variables are all $\{x\}$ except for the
two that are $\{s\}$. A possibility is $G^{\ell ,m}$, but for simplicity, we
do not pursue this issue further. One special Legendre transform is
noteworthy -- the one in which all variables are $\{s\}$. Located at the
corner of the hypercube diametrically opposite to $F$, this function will be
denoted by $H$. In this case, the Legendre transform relation simplifies to 
\begin{equation}
H(\vec{s})+F(\vec{x})=\vec{s}\cdot \vec{x}\,\,.  \label{H+F}
\end{equation}
Generalizations for higher derivatives proceed are straightforward. For
example, Eq.~(\ref{d2}) becomes 
\begin{equation}
\sum_m(\partial _\ell \partial _mH)(\partial _m\partial _nF)=\delta _{\ell
n}\,\,,
\end{equation}
where $\delta $ is the unit matrix. The convexity of $F$ guarantees that the
inverse of $\partial _m\partial _nF$ exists.

Let us apply these considerations to the thermodynamics of a gas. We begin
with the microcanonical partition function $\Omega (E,V)$ and consider the
mapping 
\begin{equation}
F(x_1,x_2)\to \mathcal{S}(E,V)\equiv \ln \Omega \,\,,
\end{equation}
$x_1\rightarrow E$, $x_2\rightarrow V$, $s_1\rightarrow \beta $, $%
s_2\rightarrow \eta $. The last of these is related to the pressure $P$, an
issue we will comment on later. The Legendre transform with respect to $x_1$
leads to the Helmholtz free energy. Our symmetric and dimensionless version
of $F=E-TS$ is same as Eq.~(\ref{F+S}): $\mathcal{F}(\beta ,V)+\mathcal{S}%
(E,V)=\beta E$, with $V$ playing the role of a ``spectator.'' Thus, to be
precise, we now write Eq. (\ref{beta}) with a partial derivative:
\begin{equation}
\beta \equiv \left. \frac{\partial \mathcal{S}}{\partial E}\right| _V\,\,.
\end{equation}
For the second Legendre transform, with respect to $x_2=V$, we define \cite
{note10} 
\begin{equation}
\eta \equiv \left. \frac{\partial \mathcal{S}}{\partial V}\right| _E
\end{equation}
and arrive at 
\begin{equation}
\mathcal{G}(\beta ,\eta )+\mathcal{S}(E,V)=\beta E+\eta V\,\,.  \label{G+S}
\end{equation}
Here, $\mathcal{G}\equiv \beta G(T,P)$ is the dimensionless Gibbs free
energy. Meanwhile, the relationship between $\eta $ and the traditional
definition of pressure, $P\equiv \left. -\partial E/\partial V\right| _S$,
is $\eta =\beta P$. To show this will take us further afield, into the first
law of thermodynamics and the notion of heat transfer. The interested reader
may consult a standard text, such as Ref. \onlinecite{note10}.

Returning to Eq.~(\ref{G+S}), we move $\mathcal{S}$ and divide both sides by 
$\beta $ to arrive at its more common form: $G=E-TS+PV$. The seemingly
mysterious signs of the last two terms on the right are, from our
perspective, due to the placing of $S$ and the use of $T$ instead of $\beta $%
. By contrast, every term comes with a positive sign in Eq.~(\ref{G+S}),
with all the potentials on the left and all the conjugate variables on the
right. Note that there are just two variables in this example, so that $%
\mathcal{G}$ plays the role of $H$ in Eq.~(\ref{H+F}), which is an explicit
writing of Eq.~(\ref{G+S}) here.

Lastly, we turn to enthalpy, which is laden with extra complications. For
various reasons, $S$ (instead of $E$) is chosen to be the independent
variable for arriving at the enthalpy. As a result, instead of $\beta $, the
natural conjugate variable is $T$ ($=\partial E/\partial S$). Regarding $S$
as a control variable with which to access $E$ is conceptually difficult.
However, it is common to think of transferring heat so that $TdS$ appears as
the means of control. If we take the Legendre transform of $E(S)$ in the
standard fashion, we would arrive at $TS-E$, which is the Helmholtz free
energy except for a sign. The disadvantage is clear, but there are
advantages to this approach. In particular, by starting with $E(S,V)$, we
naturally arrive at the ordinary pressure, $-P$, as the conjugate to $V$
(instead of $\eta $). Note the extra minus sign here. The Legendre transform
with respect to $V$ of $E(S,V)$ gives $(-P)V-E$, the (negative of) enthalpy 
$H=E+PV$. If we allow logic to overcome tradition, we would have defined the
last potential as $\mathcal{H}(E,\eta )$ (not to be confused with the
Hamiltonian $\mathcal{H}$!) through the Legendre transform 
\begin{equation}
\mathcal{H}(E,\eta )+\mathcal{S}(E,V)=\eta V\,\,,
\end{equation}
in which the first variable, $E$, plays the role of a spectator. But, the
beauty of pure reason does not always prevail and we must often abide by the
results of our historical paths.

\section{Concluding remarks}

There are many interesting aspects of the Legendre transform we have not
discussed. Covering all aspects would be more appropriate for a textbook
than a journal article. Here, let us conclude by touching on just two
important generalizations - the Legendre transform of (a) non-convex 
functions and (b) functions defined on spaces with non-trivial topology,
such as the angle on a circle - and providing references for further 
reading.

If a function is non-convex, the Legendre transform becomes multi-valued. If
we delete all but the principal branch, the Legendre transform develops
discontinuous first derivatives. If we perform another transformation, the
result would be the convex hull of the original. This topic is intimately
related to the Maxwell construction and the co-existence of phases (for
example, liquid and vapor). Although most texts on thermodynamics and
statistical mechanics discuss the Maxwell construction, few demonstrate its
relation to the Legendre transform of non-convex functions. The interested
reader may find a good example of a convexified (free energy) function in
Ref.~\onlinecite{Schroeder}.

A second generalization concerns variables whose domains have a non-trivial
topology, the simplest being functions defined on a circle or the surface of
a sphere. The angles are the most natural variables for a sphere, but we must
be mindful of the periodic nature of $\phi \in (0,2\pi ]$ and the co-ordinate
singularities at the poles $\theta =0,\pi $. An example is the shape of
crystals in equilibrium with its liquid (for example, $^4$He crystals in
coexistence with the superfluid \cite{He4}) or vapor (for example, gold
crystals\cite{Au}). Typical crystal shapes are not spherical and can be
described by a non-trivial function $R(\theta ,\phi )$, which specifies the
distance from the center of mass to a point on the crystal surface labeled
by $(\theta ,\phi )$. The tangent plane at that point can be associated by
the direction of its normal and labeled by $(\tilde{\theta},\tilde{\phi})$.
The relation between these and the derivatives $\partial _\theta R$ and 
$\partial _\phi R$ exists, but is not simple. From these derivatives a
(generalized) Legendre transform of $R$ can be constructed: $\sigma (\tilde{%
\theta},\tilde{\phi})$. The function $\sigma $ is also a significant
physical quantity: it is the free energy per unit area (surface tension)
associated with a planar interface, with normal $(\tilde{\theta},\tilde{\phi}%
)$, between the crystalline and the isotropic phases of the material. A
bonus is that, unlike typical thermodynamic potentials such as the entropy
and free energies, the ``potential'' $R(\theta ,\phi )$ is not just an
abstract concept; it can be visualized, being displayed explicitly as a
shape in three dimensions. Further details of this intriguing connection may
be found in Refs.~\onlinecite{ECS}.

Finally, we should point the readers to horizons far beyond those discussed
here. Since our purpose is to reach students and instructors in upper
undergraduate and core graduate courses, we limit our considerations to
cases with two (or finite $M$) variables above. Beyond this level, it is
possible to study the Legendre transform with an infinite number of
variables. Probably the most well known example in physics comes from both
quantum field theory\cite{QFT} and statistical field theory\cite{SFT}.
Associated with each quantum field $\phi \left( \vec{r},t\right) $ is a
``source field'' $J\left( \vec{r},t\right) $, in much the same way that a
fluctuating local magnetization, $m\left( \vec{r}\right) $, can be
``created'' by an inhomogeneous magnetic field $B\left( \vec{r}\right) $. In
the latter system, the fluctuations of $m$ are thermal, rather than quantum,
in nature. Now, the source field can be regarded as a control variable for
each $\vec{r},t$ (or just $\vec{r}$). Thus, there are an infinite number of
variables, as well as responses, involved. Corresponding to a given $J\left( 
\vec{r},t\right) $ or $B\left( \vec{r}\right) $, we can compute, in
principle, the ``vacuum energy'' $\mathcal{U}\left[ J\left( \vec{r},t\right)
\right] $ or the free energy $\mathcal{F}\left[ B\left( \vec{r}\right)
\right] $. These carry information on the quantities of interest: connected
Schwinger functions (expectation values of products of $\phi $'s) or
correlations functions (averages of products of $m$'s). More useful than 
$\mathcal{U}$ is its Legendre transform, $\Gamma $. Known as the effective
action, $\Gamma $ displays the essential information more conveniently in
terms of one particle irreducible (1PI) Schwinger functions or vertex
functions. For the effective action of a quantum field, there is a
particularly appealing systematic expansion: in powers of $\hbar $. The
zeroth order term is just the classical action. Similarly, for the Legendre
transform of $\mathcal{F}$, there is a systematic expansion in powers of $T$
or $\beta ^{-1}$. Not surprisingly, the zeroth order term here is just the
energy associated with $m\left( \vec{r}\right) $, i.e., the Hamiltonian 
$\mathcal{H}\left[ m\left( \vec{r}\right) \right] $ which enters the
Boltzmann factor $\exp \left\{ -\beta \mathcal{H}\right\} $. Our hope is
that these comments will help some students who are struggling with field
theory or perhaps further motivate those who are enthusiastically waiting to
delve into the subject.

\section{Acknowledgements}
We thank many colleagues for fruitful discussions, as well as Harvey Gould, Jan Tobochnik, and Beate Schmittmann for critical readings of the manuscript. This work is supported in part by the U.S.\ National Science Foundation through
grants DMR-0705152 and DUE-0524987.


\begin{thebibliography}{99}
\bibitem{note3}  In this example, $s$, $x$, $F$, and $G$ are all positive.
Thus, the ``$G$ axis'' points downward, opposite to the ``$F$ axis.''
\bibitem{note4}  This restriction can be lifted especially if physical quantities  with dimensions (e.g., the Hamiltonian) are to be studied. In that case, we must keep more careful track of the units, such as $[s]=[F][x^{-1}]$.
\bibitem{CUR} In general, $F$ may be regarded as a smooth $M$-dimensional manifold. The eigenvalues of $\partial _m \partial _\ell F (\vec{x})$ are the principal curvatures of this surface at $\vec{x}$.
\bibitem{note9}  These partial derivatives are taken with the understanding
that all other variables are held fixed. It is common (and reasonable) to
consider derivatives with $F$ or $G$ held fixed. In this article, we avoid
discussing such complications.
\end{thebibliography}
\end{document}